\documentstyle[11pt,aaspp]{article}

\begin{document}

\title
{A CONSISTENT EXPLANATION FOR $^{12}$C/$^{13}$C, $^7$Li AND $^3$He ANOMALIES
IN RED GIANT STARS}

\author
{C. Charbonnel}

\affil
{Observatoire Midi-Pyr\'en\'ees. LAT - CNRS URA 285. \\
14 avenue Edouard Belin. 31400 Toulouse. France}

\clearpage \newpage

\begin{abstract}
The observations of carbon isotopic ratios in evolved stars suggest that
non standard mixing is acting in low mass stars as they are ascending
the red giant branch.
We propose a simple consistent mechanism, based on the most recent
developments in the description of rotation-induced mixing by Zahn
(1992), which simultaneously accounts for the low $^{12}$C/$^{13}$C ratios
in globular cluster and field Pop II giants and for the
lithium abundances in metal-poor giant stars. It also leads to the
destruction of $^3$He produced on the main sequence in low
mass stars. This should both naturally account for the recent
measurements of $^3$He/H in galactic HII regions and allow for
high values of $^3$He observed in some planetary nebulae.
\end{abstract}


\clearpage \newpage

\section
{Evidence for non-standard mixing on the RGB
from $^{12}$C/$^{13}$C, $^7$Li and $^3$He observations}
\bigskip

At the beginning of the red giant branch (RGB), stars experience
the first dredge-up. The deepening convective envelope
brings up to the surface internal matter, which was
nuclearly-processed during the main-sequence evolution.
The dredge-up
induces in particular a decrease of the carbon isotopic ratio and of the
$^7$Li abundance, and an increase of the $^3$He abundance.
According to the standard scenario, the surface abundances then
stay unaltered as the convective envelope slowly withdraws during the
end of the RGB ascent.
However, comparisons of standard stellar evolutionary predictions
with observed abundances of different elements reveal that some
non-standard mixing mechanism is acting in low mass stars as they are
ascending the RGB.

The main evidence for this mixing comes from the behavior of the surface
carbon isotopic ratio.
In most giants with masses lower than 2 M$_{\odot}$, its value is lower
than predicted by standard evolutionary calculations (see Sneden 1991 for a
review of the problem).
Observations of evolved stars in globular and galactic clusters
indicate that the $^{12}$C/$^{13}$C ratio continues to decrease after
the completion of the theoretical first dredge-up
(Gilroy 1989; Smith \& Suntzeff 1989; Brown \& Wallerstein 1989;
Bell, Briley, \& Smith 1990; Gilroy \& Brown 1991).
This is confirmed by observations in
field population II giant stars (Sneden, Pilachowski \& VandenBerg
1986).
Observational data along the subgiant and giant branches of M67
(Gilroy \& Brown 1991) provide the major evidence that non-standard mixing
is occuring above the luminosity at which the hydrogen burning shell crosses
the chemical discontinuity created by the convective envelope during the
dredge-up (Charbonnel 1994, hereafter C94).

Observations of $^7$Li in evolved stars confirm the occurence of an
extra-mixing mechanism in low mass stars on the RGB.
In a large sample of evolved halo stars, Pilachowski, Sneden \& Booth (1993)
have shown that the lithium abundance continues to decrease after the
theoretical completion of the first dredge-up (cf Fig.2). This further decline
indicates that some mechanism transports lithium from the convective
envelope down to the region where it is destroyed by proton capture after
the end of the dilution phase.

A related problem involves $^3$He. When the $^3$He peak produced on the
main sequence is engulfed in the convective envelope of low mass stars
during the first dredge-up, the surface abundance of $^3$He increases
(cf Fig.1).
In standard stellar models,
$^3$He then survives during the following phases of evolution
(Vassiliadis \& Wood 1993), and is
injected in the ISM by stellar winds and Planetary Nebulae ejection.
In this standard view, stars of masses lower than 2 M$_{\odot}$ produce large
amount of $^3$He (Rood, Steigman \& Tinsley 1976).
$^3$He/H is thus predicted to increase with time in the regions where
stellar processing occurs.
Chemical evolutionary models including the production of $^3$He by low mass
stars
(Vangioni-Flam \& Cass\'e, Olive et al. 1995, Galli et al. 1995)
indeed predict an overproduction of $^3$He (by factors between 5 to 20)
compared to recent measurements of $^3$He/H in galactic HII regions
and in the local ISM (Rood et al. 1995). Currently observed values
essentially  leave no room to important production of $^3$He in the Galaxy,
and seem
to indicate that actually low mass stars do not produce this element.
Hogan (1995) noticed that a mixing process capable to reduce the
$^{12}$C/$^{13}$C ratios in low mass stars would also destroy $^3$He.
As we shall see, the mechanism we propose should only be efficient in
low mass stars, and it allows the high value of $^3$He/H observed in the
PN NGC 3242 (Rood, Bania \& Wilson 1992) if the initial mass of this object
was higher than 2M$_{\odot}$.

In the present paper we propose a consistent model based on a realistic
physical process,
which simultaneously accounts for observed behavior of
$^{12}$C/$^{13}$C, and $^7$Li in evolved low mass stars
and leads to low $^3$He.

\section
{Speculations on the nature of the extra-mixing process}

Different mixing processes were proposed to explain the abundance
anomalies in evolved stars.
Sweigart \& Mengel (1979) suggested that meridional circulation on the
RGB could lead to the low $^{12}$C/$^{13}$C ratios observed
in field giants.
More recently, Denissenkov \& Weiss (1995) and Wasserburg, Boothroyd \&
Sackmann (1995) reconsidered this idea in order to explain the carbon
and oxygen isotope problems on the RGB and AGB.

However, when invoking meridional circulation, one has to
consider the interaction between meridional
circulation and turbulence induced by rotation in stars, and to take
into account recent progress in the description of the transport of
chemicals and angular momentum in stellar interiors.
For this purpose,
Zahn (1992) developed a consistent theory for the mixing of chemicals
induced by rotation, in which he took the feed-back effect
due to angular momentum transport into account.
He showed that the global effect of advection moderated by horizontal
turbulence can be treated as a diffusion process.
In our context, two important points must be noted (see Zahn 1992) :
{\em 1.} The resulting mixing of chemicals
in stellar radiative regions is mainly determined by the loss of
angular momentum via a stellar wind.
{\em 2.} Additional mixing is expected near nuclear burning shells.
Since both conditions are fulfilled on the RGB, we suggest
that this process is responsible for the extra-mixing acting during the
RGB evolution of low mass stars.
In the present paper, we estimate how this rotation-induced mixing can
modify the surface values of $^{12}$C/$^{13}$C ratios and of the abundances
of $^7$Li and $^3$He.

\section
{Models}

Stellar models (0.8 and 1M$_{\odot}$ for Z=10$^{-4}$ and Z=10$^{-3}$)
are computed from the zero age main sequence up to the top
of the RGB.
We use the Geneva stellar evolutionary code
in which we have introduced the numerical method described in Charbonnel,
Vauclair \& Zahn (1992) to solve the diffusion equation.
Observations of $^{12}$C/$^{13}$C ratios in M67 evolved stars
(Gilroy \& Brown 1991) strongly suggest that the extra-mixing process
is only efficient when the
hydrogen burning shell has crossed the discontinuity in molecular weight
built by the convective envelope during the first dredge-up (C94).
Before this evolutionary point, the mean molecular weight gradient
probably acts as a barrier to the mixing in the radiative zone.
Above this point, no gradient of molecular weight exist anymore above
the hydrogen burning shell, and extra-mixing is free to act.
In our computations, we thus engage the extra-mixing on the RGB
at this evolutionary point (see Fig. 9 in C94).
Then at each numerical step we take into account both nuclear reactions
and diffusion to compute the evolution of the chemicals.
We simultaneaously treat hydrogen, lithium, and the isotopes
of carbon, nitrogen, oxygen, neon and magnesium.
The implicit method we use is extensively described in Charbonnel, Vauclair
\& Zahn (1992).

Let us stress that for stars with masses higher than 2 M$_{\odot}$
(i.e. stars which do not undergo the helium flash),
the hydrogen burning shell does not have time to reach the chemically
homogeneous region during their short ascent of the RGB.
Thus extra-mixing should naturally not occur in these more massive objects.
This crucial point is confirmed by observations in open clusters giants
(Gilroy 1989).

For the present approach, we restrict our study to the case where
the stars undergo a moderate wind on the RGB.
We make the assumption that an asymptotic regime is reached. The induced
diffusion process can then be treated with an effective coefficient of
the form  (see Zahn 1992) :
D $\simeq {{3 c_h}\over{80 \pi}} |{{dJ}\over{dt}}|
                {1 \over {\alpha \rho \Omega r^3}}$ ,
where $\alpha = {1 \over 2} {{d ln r^2 \Omega} \over {d ln r}}$,
$\Omega$ is the angular velocity, ${{dJ}\over{dt}}$ is the variation of
angular momentum, and $c_h \leq 1$. Here we take $c_h = 1$.
At each evolutionary step, D is computed in each radiative shell of the star
in order to consider its spatial and temporal evolution.
We assume a depth independent angular rotation velocity in the region
where diffusion can happen, i.e. between the base of
the deep convective envelope and the top of the hydrogen burning shell.
For all our models, we consider a constant rotation velocity of 1
km.sec$^{-1}$ on the RGB. This value is typical at this evolutionary
phase for the stellar masses we consider (De Medeiros 1990).

We use the OPAL radiative opacities (Iglesias et al. 1992) complemented
at low temperature by the atomic and molecular opacities by Kurucz
(1991). The nuclear cross-sections are from Caughlan \& Fowler (1988),
at the exception of the $^{17}$O(p,$\gamma)^{18}$F and
$^{17}$O(p,$\alpha)^{14}$N for which we adopt the values from Landr\'e
et al (1990). Screening factors are considered according to the
prescriptions by Graboske et al. (1973).
For mass loss on the red giant branch, we use the
expression by Reimers (1975). At solar metallicity, $\eta$=0.5 is
chosen (see Maeder \& Meynet 1989).
At non-solar metallicity Z, we take mass loss rates
lowered by a factor of (Z/0.02)$^{0.5}$ with respect
to the models at Z=0.020 for the same stellar parameters.

\section
{Results}

Figure 1 shows the evolution of the surface ratio $^{12}$C/$^{13}$C
as a function of luminosity,
both for standard computations and for computations including diffusion
on the RGB.
Due to dilution,
$^{12}$C/$^{13}$C decreases down to a value that
depends of the maximal extent reached by the deepening convective
envelope, and which is function of both mass and metallicity (see C94).
In the standard case the post-dilution value of the carbon
isotopic ratio stays constant for these low mass stars,
and is substantially higher than observed.
However, when extra-mixing begins to act,
the carbon isotopic ratio drops again. It reaches
the low values currently observed in globular cluster giants, i.e.
it approaches the equilibrium value in the CN-cycle, namely 3-8.
Observations in field Population II (Sneden et al. 1986)
and globular cluster M4 (Smith \& Suntzeff 1989) giant stars are
compared to the predictions in our 0.8M$_{\odot}$, Z=10$^{-4}$ model.
The theoretical slope around Log L/L$_{\odot} \simeq 2$ and the final values
obtained when considering rotation-induced mixing are in
good agreement with observations.
In a future paper, we will investigate how different rotational
histories can account for the observed dispersion.
We will also test our model from confrontation of its predictions with
observations in Population I stars, and in particular in galactic
cluster giants.

Let us now consider the problem of $^3$He.
On the main sequence, the $^3$He peak is not as deep as the
$^{13}$C peak. Thus the surface mass fraction of $^3$He begins to change
earlier in luminosity than the $^{12}$C/$^{13}$C ratio. Its value reaches a
maximum before slightly decreasing when the whole peak is engulfed in
the convective envelope.
After the dredge-up, the temperature is too low at the base
of the convective region for $^3$He to be nuclearly processed.
Then in standard models $^3$He/H stays constant,
and its final value is strongly increased compared to the initial one.
However, the temperature gradient is very steep
below the convective envelope : in our 0.8M$_{\odot}$, Z=10$^{-3}$ model,
the temperature rises from logT $\simeq$ 6.33 to $\simeq$ 7.44 in the
diffusion region at the evolutionary point where extra-mixing is supposed
to be free to act.
Thus, when $^3$He diffuses, it rapidly reaches the region where it is
nuclearly burned by the $^3$He($\alpha,\gamma)^7$Be reaction.
This leads to a rapid decrease of the surface value of $^3$He/H,
confirming the predictions by Hogan (1995).
Computations for different masses and different metallicities have now
to be performed in order to estimate the actual contribution of low mass
stars to galactic $^3$He in the framework proposed here.
Let us point out an important point concerning the high value of $^3$He/H
observed in the PN NGC 3242 (Rood et al. 1992). If confirmed,
it requires that the progenitor of this
object has not undergone extra-mixing. This can be explained
if the initial mass of this star was higher than 2M$_{\odot}$, since
extra-mixing should not develop above this stellar mass, as explained before.

Observations of the lithium abundances in Population II evolved stars
(Pilachowski et al. 1993) are displayed as a function of
effective temperature and compared with our predictions for our models
with Z=10$^{-4}$ in Figure 2.
Lithium abundances smoothly decrease with T$_{eff}$ in subgiants due to
dilution. Around 5000K the dredge-up is completed, and down to this
T$_{eff}$ the expected lithium abundance variations reproduce the
observed trend.
For stars with lower T$_{eff}$, the observed lithium abundance continues
to drop, whereas standard models do not predict further decrease.
In our models with diffusion on the RGB, lithium is rapidly
transported from the convective envelope down to the region where it is
burned by proton capture, and its surface abundance rapidly decreases
down to the very low values observed in the halo giants.
However, the problem of lithium in Population II stars
has to be considered in a more general context.
The present results will be discussed together with lithium variations
in main sequence halo stars in a forthcoming paper (Charbonnel \&
Vauclair 1995).

\section
{Summary and outlook}

The exploratory results presented here indicate that a realistic physical
process, rotation-induced mixing, can simultaneously
account for observed behavior
of carbon isotopic ratios and lithium abundances
in Population II low mass giants, and avoids large $^3$He production by
low mass stars in the Galaxy.
In a future paper, we will develop more detailed simulations in
which both the transport of matter and angular momentum will be treated
simultaneously, in order to take into account the whole rotational
history of the stars. Other chemical anomalies will be investigated.
Theoretical predictions will also be compared to observations in galactic
cluster stars. Last but not least, we will test our model efficiency in
stars on the asymptotic giant branch.

\clearpage \newpage

\centerline{FIGURE CAPTIONS}

{\em Figure1.} Evolution of $^{12}$C/$^{13}$C (top) and of $^3$He/H in units
of 10$^{-4}$ (bottom) as a function of luminosity.
Theoretical results are displayed for standard evolution (solid lines)
and for the evolution including extra-mixing (dashed-dotted lines).
As discussed in the
text, the post-dilution values remain constant in the standard case, and
depend both on mass and metallicity.
Rotation-induced mixing induces a further decrease of $^{12}$C/$^{13}$C down
to the values currently observed in Pop II giants. The final carbon
isotopic ratios are nearly identical in the three models because of their
structure similarity.
Observations of the carbon isotopic ratio in field Population II (crosses;
Sneden et al. 1986) and globular cluster M4 (
circles; Smith \& Suntzeff 1989) giant stars are compared to the
predictions of the 0.8M$_{\odot}$, Z=10$^{-4}$ model which is typical for the
considered stellar population. The behavior of $^{12}$C/$^{13}$C is well
reproduced.
Rotation-induced mixing simultaneously leads to the destruction of $^3$He
produced during the main sequence evolution. This results leads to revise
the actual contribution of low mass stars to the production of $^3$He in the
Galaxy.

{\em Figure2} Theoretical predictions for the lithium abundance variations
from the 0.8 and 1M$_{\odot}$ models computed with Z=10$^{-4}$,
compared to observations of the lithium abundances in evolved halo stars
from Pilachowski et al. (1993) (open circles for real lithium
detection, open triangles for upper limits).
The solid lines represent the lithium abundance variations due to
standard dilution alone.
The dashed lines correspond to the further decline of lithium due to
rotation-induced mixing.

\end{document}